\documentclass[11pt]{article} 
\usepackage[utf8]{inputenc}
\usepackage{multirow}
\usepackage{amsfonts}
\usepackage{amsmath}
\usepackage{amssymb}
\usepackage{amsthm}
\usepackage{caption}
\usepackage[dvipsnames]{xcolor}
    \definecolor{darkgreen}{rgb}{0,0.5,0}
    \definecolor{darkblue}{rgb}{0,0,0.6}
    \definecolor{purple}{rgb}{0.4,.2,0.7}
\usepackage[margin = 2.5cm]{geometry}
\usepackage{graphicx}
\usepackage[hyperfootnotes = false, colorlinks = true, linkcolor = darkblue, citecolor = purple]{hyperref}
\usepackage{subcaption}
\usepackage{ytableau}

\newcommand{\zmic}{Z_{\textrm{mic}}}
\usepackage{csquotes}

\usepackage{tikz}
\usetikzlibrary{decorations.pathmorphing}
\usetikzlibrary{decorations.markings}
\definecolor{mathred}{RGB}{180,44,37}
\definecolor{mathblue}{RGB}{39,94,190}
\tikzset{>=latex} % for LaTeX arrow head
\tikzset{ photon/.style={decorate, decoration={snake}, draw=black}}

\usepackage{comment}

\usepackage{physics}

\newcommand{\be}{\begin{equation}}
\newcommand{\ee}{\end{equation}}
\newcommand{\bea}{\begin{eqnarray}}
\newcommand{\eea}{\end{eqnarray}}

\def\tr{\mathrm{tr}}

\newcommand{\K}{K}
\newcommand{\sstar}{\sigma^*}
\newcommand{\acrit}{\alpha_{\textrm{crit}}}
\newcommand{\psat}{P(\textrm{SAT})}
\newcommand{\punsat}{P(\textrm{UNSAT})}

\begin{document}

\thispagestyle{empty}
\begin{center}
    ~\vspace{5mm}

  \vskip 2cm 
  
   {\LARGE \bf 
       A Jammed Parisi Ansatz  
   }

   \vspace{0.5in}
     
   {\bf Michael Winer and Aidan Herderschee
   }

    \vspace{0.5in}

  Institute for Advanced Study, Princeton, NJ 08540, USA
                
    \vspace{0.5in}

    \vspace{0.5in}

\end{center}

\vspace{0.5in}

\begin{abstract}

 Constraint Satisfaction Problems are ubiquitous in fields ranging from the physics of solids to artificial intelligence. In many cases, such systems undergo a transition when the ratio of constraints to variables reaches some value $\acrit$. Above this critical value, it is exponentially unlikely that all constraints can be mutually satisfied. We calculate the probability that constraints can all be satisfied, $\psat$, for the spherical perceptron. Traditional replica methods, such as the Parisi ansatz, fall short. We find a new ansatz, the jammed Parisi ansatz, that correctly describes the behavior of the system in this regime. With the jammed Parisi ansatz, we calculate $\psat$ for the first time and match previous computations of thresholds. We anticipate that the techniques developed here will be applicable to general constraint satisfaction problems and the identification of hidden structures in data sets.

%\AH{Mike abstract: Constraint Satisfaction Problems are ubiquitous in fields ranging from the physics of solids to artificial intelligence. In many cases, such systems undergo a transition when the ratio of constraints to variables reaches some value $\acrit$. Above this critical value, it is exponentially unlikely that all constraints can be mutually satisfied by any assignment of variables. We calculate the probability that such an assignment exists. Traditional replica methods fall short here because they capture the behavior of typical configurations, while the key properties in this regime come from exponentially rare events. We solve the problem by noting that $\psat=\lim_{n\to 0} \overline{Z^n}$ where $Z$ is the number of satisfying assignments. This limit can be calculated with a modified replica trick, using what we call the Jammed Parisi Ansatz. With this tool, we can calculate thresholds more efficiently than with previous methods, while also probing what lies beyond them.}

\end{abstract}

\vspace{1in}

\pagebreak

\setcounter{tocdepth}{3}
{\hypersetup{linkcolor=black}\tableofcontents}

\section{Introduction}\label{sec:introduction}

%\AH{add discussion about phase transition}

Constraint Satisfaction Problems (CSPs) are an important class of problems with applications ranging from physics to computer science to mathematics to artificial intelligence. The standard setup for such a problem is to have $N$ primitive variables, which can be real numbers, binary variables, or something more exotic. One has some number of $M$ of constraints on the $N$ variables, each of which must be satisfied. One can ask questions such as whether there is an assignment of the variables which satisfies all of the constraints, and what such an assignment might be. More difficult questions include counting the full number of such satisfying assignments, and sampling the space of assignments efficiently. 

The exponential size of the search space makes efficient general solutions to CSPs impossible. Boolean satisfiability, perhaps the single most famous CSP, is NP-complete. As are many other CSPs, including the ones studied in this paper. However, while solving an individual CSP is usually intractable, one valuable approach is the study of ensembles of CSPs. We can imagine drawing the constraints in our CSP from some specified distribution. Even if we cannot solve any individual instance, it is often feasible to prove statistical facts about the set of CSP with $M$ constraints on $N$ variables as $M,N\to \infty.$ 

A key quantity of interest is the probability, \(\psat\), that a given CSP instance is satisfiable for a randomly chosen element of the distribution. Another fundamental concept is the \textit{threshold}, often characterized by a critical value \(\acrit\). Typically, \(\psat\) is close to 1 when \(\frac{M}{N} < \acrit\) and close to 0 when \(\frac{M}{N} > \acrit\). Extensive research has been dedicated to determining \(\acrit\) for various CSPs, including Boolean satisfiability \cite{mezard2002analytic,M_zard_2002,achlioptas2003maximumsatisfiabilityrandomformulas}, the spherical perceptron \cite{Gardner:1987gi,Gardner1988TheSO,Franz_2016,Franz_2017}, and graph coloring \cite{frieze2003satisfiability}. In physics, these thresholds are conjectured to correspond to jamming transitions, where systems of hard grains reach a state beyond which further compression is impossible \cite{Franz_2016,Franz_2017}. Identifying the threshold is often a highly challenging problem. For instance, in the case of the spherical perceptron, it is believed that a full replica-symmetry-breaking calculation is required. 

However, to our knowledge, very little has been done to calculate $\psat$ for almost any type of problem.\footnote{Ref. \cite{cruciani2022capacityneuralnetworks} is a notable exception.} There are two motivations for such a calculation: threshold detection as above and structure detection in data sets. To study thresholds, one could calculate $\acrit$ by seeing when instability develops in the $\psat$ computation. Regarding structure detection, Ref. \cite{Gardner:1987gi} was the first to show how memorizing training data can be seen as a CSP. However, in real-life applications, the constraints are far from random, often leading to a vast increase in learnability \cite{mignacco2024nonlinearclassificationneuralmanifolds,Chou2024.02.26.582157,PhysRevLett.131.027301,chung2016linear,cohen2020separability,chung2018classification}. $\psat$ indicates the probability that an architecture could learn unstructured data. If $\psat<1$, we know that this data is unusually structured and that less than $\psat$ fraction of possible datasets is more learnable. One can then propose a structure for the data (say that the points all lie on a collection of low-dimensional manifolds) and calculate a new $\psat$ using these new constraints. If the new $\psat$ is still less than one, some additional structure is still to be understood.

To calculate $\psat$ for a system of constraints, we make use of the \textit{Gardner volume} $Z$, the size of the solution space for a specific choice of the constraints. Since 
\begin{equation}
    \lim_{n\to 0}Z^n=\begin{cases}0&Z=0\\ 1&Z>0\end{cases}
\end{equation}
we have the identity
\begin{equation}
    \psat=\lim_{n\to 0} \overline{Z^n} \ .
    \label{eq:pZ}
\end{equation}
Traditional replica calculations focus on systems like the $p$-spherical model or Sherrington-Kirkpatrick model where the limit $\lim_{n\to 0}\overline{Z^n} $is identically 1, and the simplest interesting quantity is the free energy $\lim_{n\to 0} \partial_n \overline{Z^n}$. In this case, one can rely on calculations based on the Parisi Ansatz, described in \cite{infiniteOrder,parisi1980,parisi1983,mezard1987,talagrand2006,panchenko2013}. Unfortunately, the Parisi Ansatz will fail in cases where $Z$ is often zero, because it cannot give a value of $\overline{Z^n}$ which is different from one. In cases like the overconstrained CSP, this manifests as an instability where $\partial_n \overline{Z^n}$ diverges. However, the jammed Parisi ansatz (JPA) is a generalized form of overlap matrix whose entries are allowed to depend on $n$. JPA overlap matrices are chosen so that $\lim_{n\to 0}\overline{Z^n}$ can be different from 1. For the replica symmetric JPA (RS JPA), this takes the form
\begin{equation}
    Q_{ij}=\gamma n(\delta_{ij}-1)+1.
\end{equation}
where $Q_{ij}$ is the overlap matrix. Optimizing over JPA matrices, one can calculate $\psat$ for a wide variety of CSP. In fact, one can often get extremely precise values of $\psat$ with just the RS JPA.

Much of this note will focus on one of the simplest CSP: the \textit{spherical perceptron}. Our constraints are a system of linear inequalities: 
\begin{equation}
\begin{split}
\vec{\xi}_\mu\cdot \vec{x}>\sigma,\\ 
|\vec{x}|=\sqrt N
\label{eq:constraints}
\end{split}
\end{equation}
where $\vec{x}$ is an $N$-dimensional vector to be determined and $\xi_\mu$ is a fixed $N-$dimensional vector for each $\mu=1,\dots,M$. We will assume that each element of each $\xi$ is drawn from Gaussian distributions with variance $\frac{1}{N}$, so that each $\xi_\mu$ is approximately a unit vector. The model has a long history, having been studied first in an artificial intelligence context, and also as a model of jamming transitions in physics \cite{Franz_2016,Artiaco_2021}. Ref. \cite{Gardner:1987gi} showed that in the $\sigma=0$ case, there is a phase transition when the ratio $\alpha=M/N$ reaches $\acrit=2$. For $\alpha>2$, the system is over-constrained and $\psat$ is small. As $M$ decreases below $2N$, the system becomes under-constrained, and $\punsat=1-\psat$ is small. We will compute the threshold of this transition, $\acrit$, as a function of $\sigma$ along with $\psat$ using the JPA. For positive $\sigma$, the system is replica symmetric and our results do not differ from earlier work. For negative $\sigma$, where the system exhibits full replica symmetry breaking (RSB), the replica symmetric JPA still allows us to compute $\acrit$ and $\psat$ to a high degree of precision. 

One could calculate $\acrit$ by computing $\lim_{n\to 0} \partial_n \overline{Z^n}$ using a Parisi ansatz, and seeing when the instability develops. We perform this calculation up to 2RSB, showing that it agrees exactly with the 1RBS Jammed Parisi Ansatz. However, the calculation using any level of the Parisi ansatz does not give $\psat$. 

The plan for the paper is as follows. Section \ref{sec:review} is an overview of traditional replica methods as applied to the spherical perceptron. We derive the replicated action and get an expression for the satisfiability threshold in the replica symmetric cast.

Section \ref{sec:JPA} is the main technical meat of the paper: we derive the action for the spherical perceptron in the Jammed Parisi Ansatz, in both the replica symmetric and RSB cases. We calculate $\psat$, showing that even in the RS JPA it is a much lower threshold than the RS Parisi ansatz would predict, and examine the second-order nature of the transition.

Finally, in section \ref{appendixgroundstatenergypspherical} we consider another application of the JPA: the computation of ground-state energies. As a proof of concept, we compute the ground-state energy of the $p$-spherical model. To the authors' knowledge, the JPA ansatz is arguably the most efficient method for computing the ground state energies of disordered systems.\footnote{Note that this computational strategy would not catch the failure of large-$N$ perturbation theory described in Ref. \cite{Herderschee:2024zlc}.}

\section{The Parisi Ansatz in the Spherical Perceptron}
\label{sec:review}

The goal of this section is to review the standard approach to calculating $\acrit$ for this system. This involves interpreting the size of solution space as a partition function, and then using the famous replica trick to calculate the free energy. In this section, we will review the setup of the calculation in terms of the replica trick. We will not perform a RSB analysis here; instead, we refer the reader to \cite{Franz_2017} for details.

\subsection{Converting to the overlap matrix}

The key quantity in our analysis of the linear system is 
\begin{equation}
    Z=\int_{S^N} d^{N}\vec{x} \prod_{\mu=1}^M\theta(\vec{\xi}_\mu \cdot \vec{x}-\sigma) .
\end{equation}
This quantity the volume of phase space that satisfies the $M$ constraints, is often called the Gardner volume. $Z$ is a random variable that depends on all $M\times N$ variables $\vec{\xi_{\mu}}$ which are drawn from a Gaussian ensemble 
\begin{equation}\label{gaussiandistro}
P(\vec{\xi})=\frac{1}{\sqrt{2\pi \sigma^{2}}}e^{-\frac{\vec{\xi}^{2}}{2\sigma^{2}}}|_{\sigma=1/\sqrt{N}} \ .
\end{equation}
If $Z=0$, the system of inequalities is unsatisfiable. If $Z$ is positive, the system of inequalities is satisfiable. The size of $Z$ measures the size of the solution space for a given set of $\xi$s. $\overline{Z}$, the average value over the $\vec{\xi}$s, gives the average size of solution space.

More important than $\overline Z$ is $\overline {\log Z}$. When there is a large probability of $Z$ being 0, this expectation becomes infinite. Thus, $\overline {\log Z}$ gives us a window into the transition to unsatisfiability. We use the replica trick to compute $\overline {\log Z}$.\footnote{Note that $\overline {\log Z}\neq \log \overline {Z}$.} The philosophy behind this trick is explained extensively in the literature, see Refs. \cite{Castellani_2005,Mzard1987SpinGT,Fischer_Hertz_1991,De_Dominicis_Giardina_2006}. We will attempt to give only a brief review. The fundamental idea is that the $n$th power of the partition function $Z^n$ can be thought of as the partition function of a supersystem with $n$ copies of the original system. One calculates the expected value of this replicated partition function for general $n$, giving $\overline{Z^n}$. Then, if one is interested in $\overline{\log Z}$, one makes use of $\log Z=\partial_n Z^n|_{n=0}$, which we compute via saddle approximation in the large-$N$ limit. To find the critical threshold, we find the critical ratio $\acrit$ at which there is no valid saddle for $\overline{\log Z}$. Again, note that this method does not provide $\psat$ when $M/N>\acrit$, but instead just computes the phase transition point $\acrit$. 

To see this transition more explicitly, we compute $\overline{Z^{n}}$ as a function of the overlap matrix. We first note that the explicit ensemble averaged expression for $\overline{Z^{n}}$ is 
\begin{equation}
\begin{split}
\overline{Z^{n}}= \int \prod_{i}^{n} d^{N}\vec{x}_{i}\delta(\vec{x}_{i}^{2}-N)\prod_{\mu}^{M}d^{N}\vec{\xi}_{\mu} P(\vec{\xi}_{\mu})\theta(\vec{\xi}_{\mu}\cdot \vec{x}_{i}-\sigma)
\end{split}
\end{equation}
where $P(\vec{\xi})$ is given in Eq. (\ref{gaussiandistro}). For each $\vec{\xi}_{\mu}$ integral, we perform a change of variables. 
\begin{equation}
\vec{\xi}=(h_{1},\ldots,h_{n},\vec{\xi}_{\perp}), \quad \textrm{where}\quad h_{i}=\vec{\xi}\cdot \vec{x}_{i}, \quad \vec{x}_{i}\cdot \vec{\xi}_{\perp}=0
\end{equation}
and integrate over $\vec{\xi}_{\perp}$. Accounting for the Jacobian factor, the result is $M$ products of 
\begin{equation}
\begin{split}
F(\vec{x}_{i}\cdot \vec{x}_{j})&=\int d^{N}\vec{\xi}_{\mu} P(\vec{\xi}_{\mu})\prod_{i}^{n}\theta(\vec{\xi}_{\mu}\cdot \vec{x}_{i}-\sigma)\\
&=\frac{1}{\sqrt{\det 2\pi (x_{j}\cdot x_{i})}}\int d^{n}h_{i} e^{-\frac{1}{2}h_{i}(x_{j}\cdot x_{i})^{-1}h_{j}}\prod_{i=1}^{n} \theta(h_{i}-\sigma) \ .
\end{split}
\end{equation}
We then insert the overlap matrix $Q_{ij}=\vec{x}_{i}\cdot \vec{x}_{j}$ into the original integral
\begin{equation}
1=\int d^{n\times n}Q_{ij}\delta(NQ_{ij}-\vec{x}_{i}\cdot \vec{x}_{j})=\frac{1}{(4\pi)^{\frac{n(n+1)}{2}}}\int d^{n \times n}\Sigma_{ij}d^{n\times n}Q_{ij} e^{\frac{1}{2}\sum_{i\leq j}\Sigma_{ij}(NQ_{ij}-\vec{x}_{i}\cdot \vec{x}_{j})}
\end{equation}
exchange the delta function enforcing the spherical constraint with 
\begin{equation}
\prod_{i}\delta(\vec{x}_{i}^{2}-N)=\frac{1}{(2\pi)^{n}}\int d\lambda_{i}e^{\lambda_{i}(\vec{x}_{i}^{2}-N)}
\end{equation}
and finally integrate over $\vec{x}_{i}$. The end result is 
\begin{align}\label{eq:fullAction}
\overline{Z^n}&=C\int d^{n\times n}Qd^{n\times n}\Sigma d^n\lambda \exp\bigg(M\log F(Q)-\frac 12 N\log \Sigma+\frac 12 N\sum_{i\leq j}\Sigma^{ij}Q_{ij}+N\sum_{i}\lambda_i(Q_{ii}-1)\bigg) \nonumber \\
&\textrm{where}\quad C=\frac{(2 \pi )^{-\frac{1}{2} n (n-N+3)}}{2^\frac{n(n+1)}{2}}\ . 
\end{align}
We are left with an integral in terms of $Q$, $\Sigma$, and $\lambda$. The contours for $\Sigma$ and $\lambda$ need to be the imaginary axis, perpendicular to the real line. Noticing that every term in the exponent is proportional to $N$, we can use the saddle-point method to solve Eq. (\ref{eq:fullAction}). We will assume the saddle point formula is valid even in the $n\rightarrow 0$ limit. Importantly, one usually tries to find the $Q$ which maximizes the log of the partition function, but the steepest descent contour actually minimizes the log for $n<1$ \cite{Castellani_2005,Mzard1987SpinGT,Derrida_2015}. Applying the saddle-point approximation at large-$N$, we solve the saddle point equations for $\Sigma$ and $\lambda$, ultimately finding
\begin{equation}
\begin{split}\label{tominimcollectivecoord}
\lim_{N\rightarrow \infty}\frac{\log(\overline{Z^{n}})}{N}&= \begin{cases}
\max_{Q} \alpha \log F(Q)+\frac{1}{2}\tr \log(Q)+\frac{n}{2}+\frac{n}{2}\log(2\pi)& \text{if}\;n\geq 1 \\
\min_{Q} \alpha \log F(Q)+\frac{1}{2}\tr \log(Q)+\frac{n}{2}+\frac{n}{2}\log(2\pi)& \text{if}\;n<1 
\end{cases} \\
&\textrm{ where } \forall_{i}:\quad  Q_{ii}=1
\end{split}
\end{equation}
We can now compute when there is a valid saddle approximation of $\overline{Z^{n}}$ to study $\acrit$. For the remainder of the paper, we will assume we are working in the $N\rightarrow \infty$ limit and suppress the $\lim_{N\rightarrow \infty}$.  

\subsection{Replica symmetric ansatz}\label{sec:rsansatz}

As an explanatory example, let us consider a replica symmetric ansatz compatible with the spherical constraint, 
\begin{equation}
\forall 1\leq i,j\leq n: \quad Q_{ij}=(1-q)\delta_{ij}+q
\end{equation}
which dominates for $\sigma>0$. Every term in Eq. (\ref{tominimcollectivecoord}) is trivial to compute except $F(Q)$. To evaluate $F(Q)$, we add an auxiliary integration variable, $h_{0}$ and extend $Q$ to 
\begin{equation}
\forall i,j,\quad  0\leq i,j\leq n: \quad Q_{ij}'=c\delta_{ij}(1+\delta_{i0}q)+q
\end{equation}
such that 
\begin{equation}
\int \frac{d^{n}h_{i}}{\sqrt{\det 2\pi Q}} e^{-\frac{1}{2}\sum_{i,j}h_{i} (Q_{ij}^{-1})h_{j}}\prod_{i=1}^{n} \theta(h_{i}-\sigma)=
\int \frac{d^{n+1}h_{i}}{\sqrt{\det 2\pi Q'}} e^{-\frac{1}{2}\sum_{i,j}h_{i} (Q_{ij}'^{-1})h_{j}}\prod_{i=1}^{n} \theta(h_{i}-\sigma)
\end{equation}
and make the change of variables 
\begin{equation}
\forall i,\ 1\leq i\leq n: \quad h_{i}=\delta h_{i}+h_{0} \ .
\end{equation}
The resulting integral drastically simplifies
\begin{equation}
\begin{split}
    F(Q)=\int \exp\left(-\frac{h_0^2}{2q}\right)\Theta\left(\frac{h_0-\sigma}{\sqrt{1-q}}\right)^n\frac{dh_0}{\sqrt{2\pi q}}
\end{split}
\end{equation}
where we define
\begin{equation}
    \Theta(z)=\int_{-\infty}^{z} \exp\left(-\frac{h^2}{2}\right)\frac{dh}{\sqrt{2\pi}}
\end{equation}
for convenience. In the limit $n\to 0$, we have 
\begin{equation}\label{Fintegralclosen1}
\begin{split}
    \lim_{n
    \rightarrow 0}\log F= n\int \exp\left(-\frac{h_0^2}{2q}\right)\log \Theta\left(\frac{h_0-\sigma}{\sqrt{1-q}}\right)\frac{dh_0}{\sqrt{2\pi q}}\end{split}
\end{equation}
This is essentially a Gaussian integral of the log of the error function. 

%There is no way to evaluate it analytically for general $q$.

We do not care about computing $F(Q)$ for generic $\alpha$, but instead near threshold. For $\alpha>\acrit$ for some $\acrit$, Eq. (\ref{tominimcollectivecoord}) becomes arbitrarily negative as $q\rightarrow 1$ and has no saddle. This means that $\frac 1n \log \overline{Z^n}$, the expected value of $\log Z$, now has an expectation value of negative infinity. This is precisely our phase transition, beyond which $Z$ is zero with high probability. We wish to compute $\acrit$ as an explicit function of $\sigma$. We make the approximation $q\sim 1$, which is valid for $\alpha \sim \acrit$, and evaluate Eq. (\ref{Fintegralclosen1}) analytically. We see that 
\begin{equation}
    \log \Theta\left(\frac{h_0-\sigma}{\sqrt{1-q}}\right)\approx 
    \begin{cases}
	-\frac{(h_0-\sigma)^2}{2(1-q)}, & \text{if }h_0-\sigma<0\\
        0, & \text{if }h_0-\sigma>0
    \end{cases}
\end{equation}
so
\begin{equation}
    \log F\approx -\frac{1}{2(1-q)}\int_{-\infty}^\sigma \exp\left(-\frac{h_0^2}{2}\right)(\sigma-h_0)^2\frac{dh_0}{\sqrt{2\pi}}=-\frac{\K_1(\sigma)}{2(1-q)},
\end{equation}
where we define 
\begin{equation}
    \K_1(\sigma)=\int_{-\infty}^\sigma \exp\left(-\frac{h_0^2}{2}\right)(\sigma-h_0)^2\frac{dh_0}{\sqrt{2\pi}} \ .
\end{equation}
Plugging this back into our action, we have
\begin{equation}
    \frac 1N \overline{\log Z}=\frac{1}{nN}\log\overline{Z^n}\approx \min_q\left(-\alpha \frac{\K_1(\sigma)}{2(1-q)}+\frac 12 \log(1-q)+\frac 12 \left ( \frac{q}{1-q}\right )-\frac 12\right) \ .
    \label{eq:jamming}
\end{equation}
We see that there is a phase transition at 
\begin{equation}\label{acritsigmag0}
\acrit=\frac{1}{K_{1}(\sigma)} \ .
\end{equation}
Whenever $\alpha$ is above this threshold, Eq. (\ref{eq:jamming}) has no local minimum, becoming arbitrarily negative as $q$ approaches one. 

\subsection{Beyond replica symmetry}\label{brspreview}
\begin{figure}
    \centering
    \includegraphics[width=0.2\linewidth]{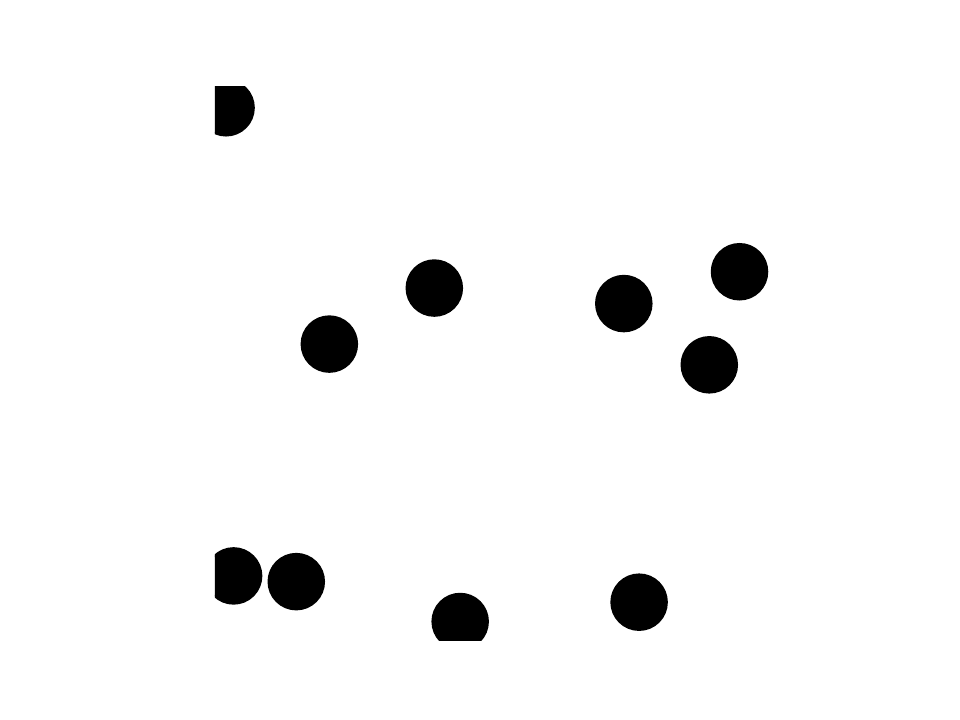}
    \includegraphics[width=0.2\linewidth]{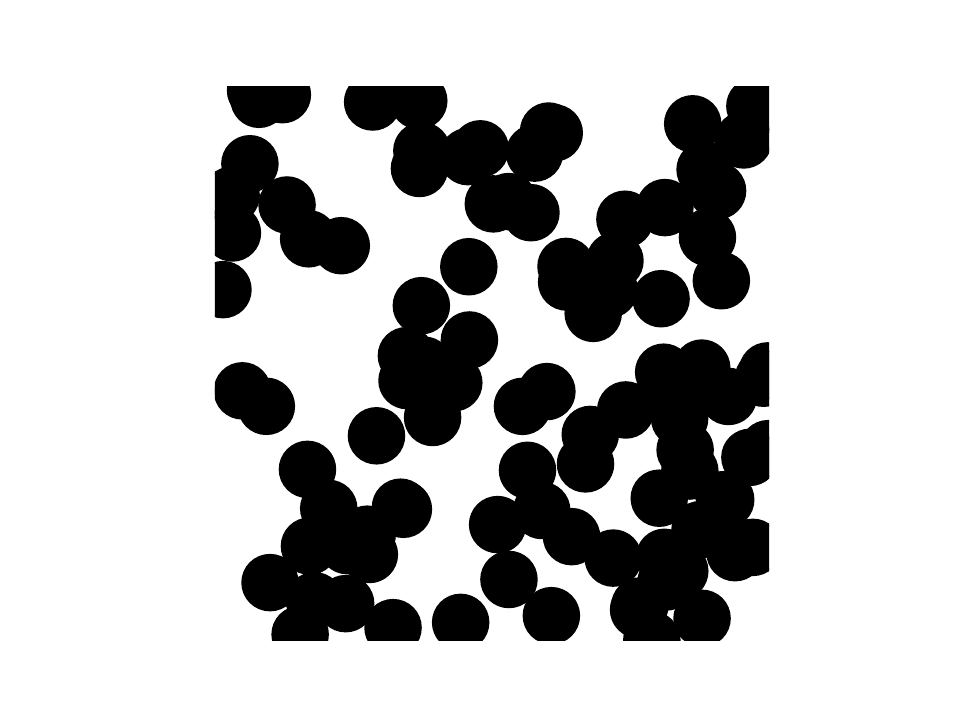}\\
    \includegraphics[width=0.2\linewidth]{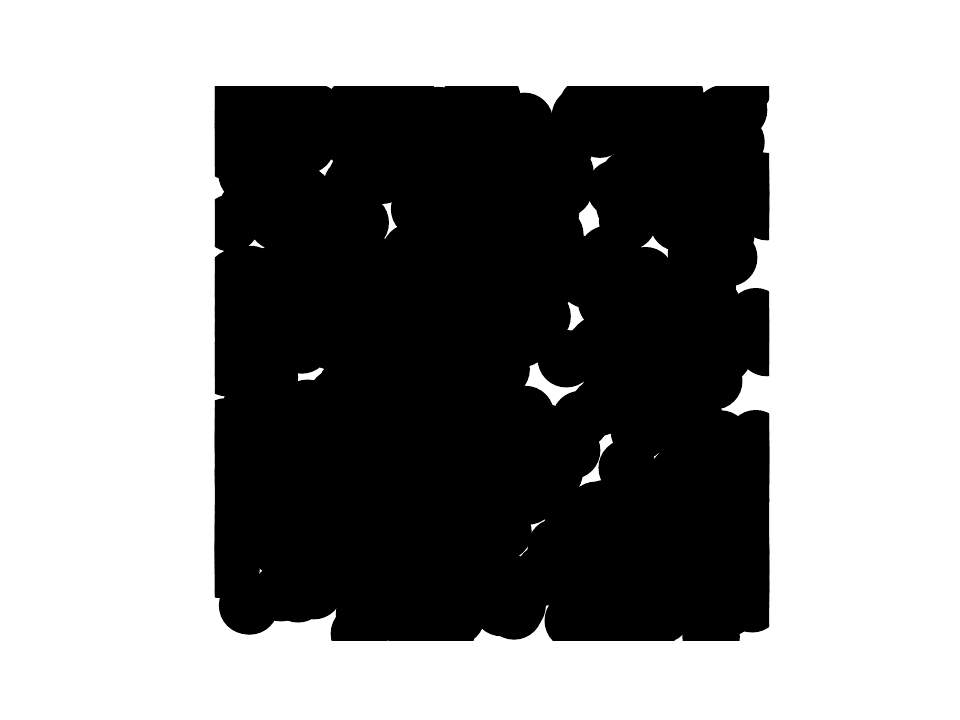}
    \includegraphics[width=0.2\linewidth]{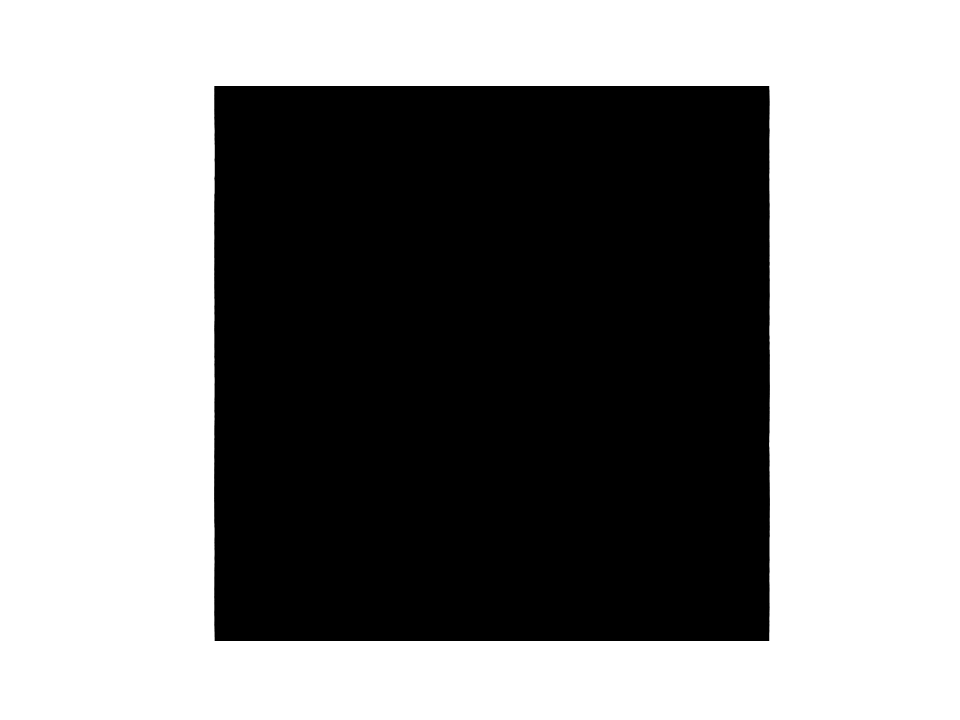}
    \caption{Each circle eliminates a small region of phase space, analogous to each constraint when $\sigma$ is extremely negative. At low densities, the white region is connected. At higher densities, it shatters, until finally the black constraints cover every point.}
    \label{fig:4densities}
\end{figure}

The RS Parisi ansatz in Eq. (\ref{acritsigmag0}) is only valid for $\sigma>0$. For $\sigma<0$, one needs to use an ansatz that explicitly breaks replica symmetry. One important generalization is the $k$th-order RSB ($k$RSB) Parisi ansatz. In the $k$RSB ansatz, the $n\times n$ matrix is made out of blocks of size $m_0$ which is made out of blocks of size $m_1$ all the way down to blocks of size $m_k=1$. Although we will not work through the details here, one can calculate both $\det Q$ and $F(Q)$ for this ansatz.

In order to build physical intuition, consider a 1RSB system whose saddle-point overlap matrix, $Q$, has some value of $m$ between 0 and 1. The physical interpretation of this overlap matrix is that the allowed region of phase space has broken into disconnected blobs. Replicas in the same blob have overlap $q_1$, replicas in different blobs have overlap $q_0<q_1$.

For $\sigma>0$, the region of the sphere satisfying all the constraints must always be simply connected and convex. This implies single region of phase space corresponds to a replica-symmetric saddle point for $Q$. By contrast, when $\sigma<0$, each constraint cuts out just a small fraction of phase space (a share of about $e^{-\sigma^2/2}$). After enough small portions of phase space are removed, it becomes disconnected into multiple components. The components are blotted out one by one, and only then does the CSP become unsatisfiable. For negative $\sigma$, we expect the true saddles to display RSB.

This explains why we expect the RS ansatz in Section \ref{sec:rsansatz} to fail. Since we are only analyzing a subset of $Q$s (the replica symmetric ones), we should expect the true value of $\overline{Z^n}$ to be lower, and thus the Gardner volume should go to zero at an even lower value of $\alpha$. This extremely complicated analysis was carried out in \cite{annesi2025exactfullrsbsatunsattransition}, and the authors got a value for $\acrit$ identical to ours. 

\section{The Jammed Parisi Ansatz}
\label{sec:JPA}
The previous section focused on computing $\overline{\log Z}$. However, we want to compute $\psat$ directly, which is given by 
\begin{equation}\label{psatformulass}
\psat=\lim_{n\rightarrow 0}\overline{Z^{n}} \ .
\end{equation}
This probability might be close to 1 (if $\alpha<\acrit$) or exponentially small (if $\alpha>\acrit$). It is this phase transition, and what lies beyond it, that we will investigate in this section. 

Interestingly, we find that the usual Parisi ansatz is insufficient. Instead, we must use a generalization of the Parisi ansatz, the JPA, where elements of $Q_{ij}$ explicitly depend on $n$. We will give the RS JPA in Section \ref{replicasymJPA}. In Section \ref{RSBJPA}, we give a $k$th-order replica symmetry breaking JPA ($k$RSB JPA) that is analogous to the $k$RSB Parisi ansatz. We believe the full JPA is necessary to reproduce the exact $\acrit$ and $\psat$ for $\sigma<0$, at least for the spherical perceptron. However, we find even the RS JPA is sufficient to get a very accurate result.

\subsection{The replica symmetric JPA}\label{replicasymJPA}

\begin{figure}
    \centering
    \includegraphics[width=0.6\linewidth]{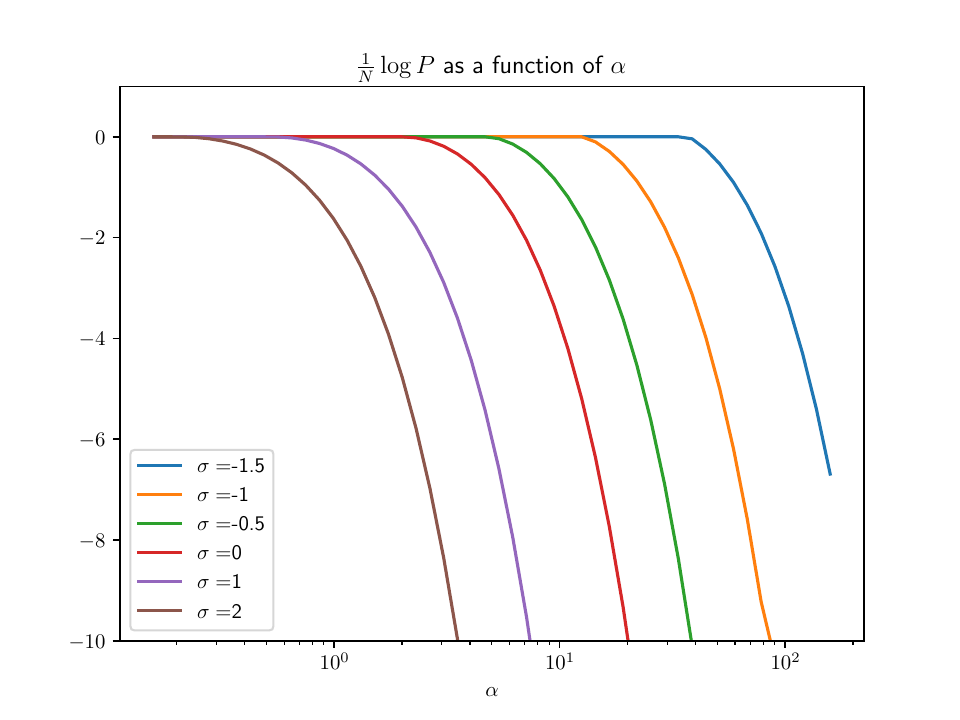}
    \caption{A graph of $\frac 1N \log \psat$ as a function of $\alpha$ for various choices of $\sigma$. For larger values of $\sigma$, the transition appears to be second order. For the most negative values, one sees a discontinuity in the derivative at the transition point.}
    \label{fig:logPGraph}
\end{figure}

We showed in section \ref{sec:rsansatz} that there is no fixed saddle point value of $q$ in the $n\to 0$ limit for $\alpha>1/\K_1(\sigma)$. Instead, the action increases as $q$ approaches one. This motivates us to introduce a new ansatz: $q=1-\gamma n$. More explicitly, the RS JPA is given by 
\begin{equation}
\textrm{RS JPA}: \quad Q_{ij}=\gamma n\left(\delta_{ij}-1\right)+1 \ .
\end{equation}
The calculation is formally the same as Section \ref{sec:rsansatz}, except that there is additional $n$ dependence. In particular, 
\begin{equation}
\lim_{n\rightarrow 0}F(Q)=\log \int \frac{dh_{0}}{\sqrt{2\pi}}\exp\left ( \frac{-h_{0}^{2}}{2}\right )\exp \left ( -\frac{\min(h_{0}-\sigma,0)^{2}}{2\gamma}\right ) \ .
\end{equation}
In the JPA, $\log \lim_{n\to 0} \overline{Z^n}$ has a fixed limit, not one proportional to $n$ as with most thermodynamic systems. This limit is given by
\begin{equation}\label{tobeminimized}
\begin{split}
    \frac 1N\log \psat&=\min_{\gamma}\left(\frac 12 \log \left(\frac{\gamma+1}{\gamma}\right)+\alpha \log \int \exp\left(-\frac{h_0^2}{2}\right)\exp\left(-\frac{\min(h_0-\sigma,0)^2}{2\gamma}\right)\frac{dh_0}{\sqrt{2\pi}}\right)
    \end{split}
\end{equation}
We can minimize Eq. (\ref{tobeminimized}) numerically. Results are plotted in Fig. \ref{fig:logPGraph}. The first term in Eq. (\ref{tobeminimized}) is always positive and decreases towards 0 as $\gamma$ increases. The integral term is always negative and increases towards 0. For small $\alpha$, Eq. (\ref{tobeminimized}) is minimized by taking $\gamma\to\infty$ which sets Eq. (\ref{tobeminimized}) to zero, consistent with our finding in Section \ref{sec:rsansatz}.

\begin{figure}
    \centering
    \includegraphics[width=0.6\linewidth]{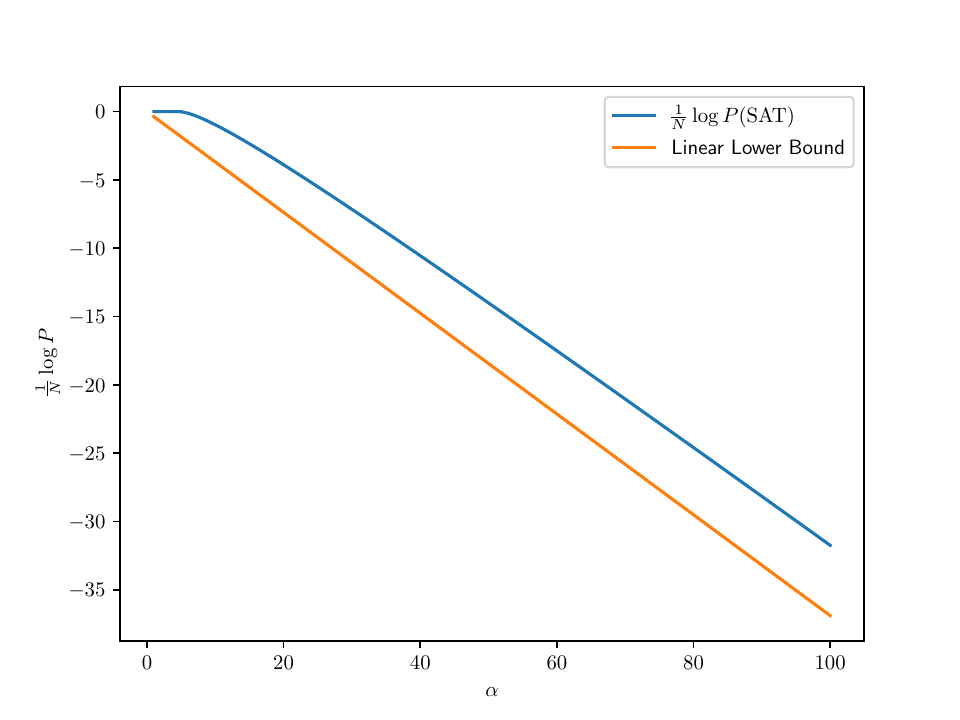}
    \caption{A plot with $\sigma=-0.5$ illustrating the true value of $\log \psat$ compared to the lower bound in equation \ref{eq:lowerBound}. We see that it does indeed have the correct asymptotics for large $\alpha$.}
    \label{fig:lowerBound}
\end{figure}

For larger $\alpha$, the integral term will dominate, forcing a negative value of Eq. (\ref{tobeminimized}) at a finite value of $\gamma$. We can attempt a lower bound by noticing that $\log \left(\frac{\gamma+1}{\gamma}\right)$ is always positive and $\log \int \exp\left(-\frac{h_0^2}{2}\right)\exp\left(-\frac{\min(h_0-\sigma,0)^2}{2\gamma}\right)\frac{dh_0}{\sqrt{2\pi}}$ is always greater than $\log \int_\sigma^\infty \exp\left(-\frac{h_0^2}{2}\right)\frac{dh_0}{\sqrt{2\pi}}$, a negative number. Thus we have
\begin{equation}
    \frac 1N \log \psat \geq \alpha \log \int_\sigma^\infty \exp\left(-\frac{h_0^2}{2}\right)\frac{dh_0}{\sqrt{2\pi}}
    \label{eq:lowerBound}
\end{equation}
This lower bound corresponds exactly to the probability that any given point satisfies all the constraints. In other words, the probability that a CSP is satisfiable is lower-bounded by the probability that a given point satisfies all the constraints. This lower bound is a good approximation of $\log \psat$ at large $\alpha$, as illustrated in Fig. \ref{fig:lowerBound}. Even though this ansatz is replica symmetric, it gives different, seemingly more accurate answers than the vanilla RS calculation of the partition function we saw in Section \ref{sec:rsansatz}.

One could wonder why we did not consider a $q$ with more general $n$ dependence. Let us assume a slightly more general ansatz 
\begin{equation}
q=1-\gamma n^{a}
\end{equation}
where $a>0$. For $a<1$, the $N\log(Q)$ in Eq. (\ref{tominimcollectivecoord}) vanishes in the limit that $n\rightarrow 0$ while $F(Q)$ is finite. Therefore, if there were a saddle, it would be independent of $\alpha$, which is unphysical. In the case where $a>1$, the $\log(Q)$ term diverges while the $F(Q)$ term is again finite, leading again to an unphysical result. Therefore, we are naturally led to $a=1$.

Let us consider the physical interpretation of $\gamma$. For an allowed region of the $\vec{x}_{i}$ phase space, the characteristic length scale is 
\begin{equation}
i\neq j:\quad \overline{(\vec{x}_{i}-\vec{x}_{j})^{2}} =2 \gamma n  \ .
\end{equation}
Therefore, $\gamma$ parametrizes how quickly the size of such regions goes to zero as $n$ goes to zero. 

\subsection{Replica symmetry breaking JPA}\label{RSBJPA}

The RS JPA has saddle points in the $n\to 0$ limit even in the unsatisfiable phase. Importantly, these saddle points have an action which remains constant as $n\to 0$. There are, however, more general ansatzes with this property. One, which inherits the structure of the $k$RSB Parisi ansatz described in Section \ref{brspreview}, is the following. 

The overlaps matrix consists of blocks of size $m_{k}=1$ inside blocks of size $m_{k-1}<1$ inside blocks of size $m_{k-2}<m_{k-1}$ all the way until a block of size $m_{-1}=n$. $m_i$ is chosen to scale with $n$ for $i<k$, going as $m_i=\tilde m_i n$. If a matrix element is inside a diagonal block of size $m_i$ but not a diagonal block of size $m_{i+1}$, that element is $q_{i+1}$. We will choose all $q$s to be fixed as $n\to 0$ except $q_k$, which goes as $1-\gamma n$. The overlap distribution will be $1-\gamma n$ with probability $\frac{1-\tilde m_{k-1}n}{1-n}$, which goes to $1$ as $n$ goes to 0. Thus, even though we have replica symmetry breaking, the chance that any two replicas are in the same minimum is 1, consistent with the idea that in the rare satisfiable cases in the over constrained regime there is likely to be only one patch of phase space which satisfies all the constraints.

\begin{table}
\centering
\begin{tabular}{c|c}
         Eigenvalues& Degeneracy \\
         \hline
         $\gamma n$&  $n-\frac 1{\tilde m_{k-1}}$\\
         $\left(\gamma+(1-q_{k-1})\tilde m_{k-1}\right)n$&$\frac 1{\tilde m_{k-1}}-\frac 1{\tilde m_{k-2}}$\\
         $\vdots$&$\vdots$\\
         $\left(\gamma+\sum_{i=0}^{k-1}(q_{i+1}-q_{i})\tilde m_{i}\right)n$ &$1$
    \end{tabular}
\caption{Eigenvalues of the RSB JPA.}\label{tab:my_label}
\end{table}

We can perform a calculation analogous to Section \ref{replicasymJPA}. The calculation of the log-determinant is fairly straightforward. The degeneracies of $Q$ are given in Table \ref{tab:my_label}. The calculation of the $\log(F(Q))$ is more cumbersome. However, we can repeat a similar trick as in Section \ref{sec:rsansatz} to find a recursive expression for $F(Q)$
\begin{equation}
\begin{split}
    g_{k}(h)&=\exp\left(-\frac{\textrm{min}(h-\sigma,0)^2}{2\gamma}\right)\\
    g_{i}(h)&=\int_{-\infty}^\infty g_{i+1}(h')^{\frac{\tilde m_{i}}{\tilde m_{i+1}}}\exp\left(\frac{(h-h')^2}{2(q_{i+1}-q_i)}\right)\frac{dh'}{\sqrt{2\pi(q_{i+1}-q_{i})}}\\
    g_{-1}(h)&=\int_{-\infty}^\infty g_0(h')^{\frac{1}{\tilde m_{0}}}\exp\left(\frac{(h-h')^2}{2q_0}\right)\frac{dh'}{\sqrt{2\pi q_0}}.
\end{split}
\end{equation}
where $F(Q)=g_{-1}(0)$, $\tilde{m}_{k}=1$, and $q_{k}=1$.

\begin{table}
    \centering
    \begin{tabular}{rccc}
        \hline
        $\sigma$ & RS PA & RS JPA & 1RSB JPA \\
        \hline
        -3.0 & 4915.573 & 2526.293 & 2526.143 \\
        -2.9 & 3390.300 & 1803.603 & 1803.416 \\
        -2.8 & 2358.003 & 1298.765 & 1298.540 \\
        -2.7 & 1653.743 & 943.216  & 942.944  \\
        -2.6 & 1169.455 & 690.775  & 690.447  \\
        -2.5 & 833.804  & 510.098  & 509.707  \\
        -2.4 & 599.351  & 379.757  & 379.295  \\
        -2.3 & 434.317  & 284.989  & 284.453  \\
        -2.2 & 317.256  & 215.553  & 214.940  \\
        -2.1 & 233.592  & 164.287  & 163.601  \\
        -2.0 & 173.348  & 126.152  & 125.402  \\
        -1.9 & 129.646  & 97.573   & 96.775   \\
        -1.8 & 97.711   & 76.001   & 75.173   \\
        -1.7 & 74.205   & 59.599   & 58.763   \\
        -1.6 & 56.779   & 47.041   & 46.218   \\
        -1.5 & 43.769   & 37.360   & 36.567   \\
        -1.4 & 33.989   & 29.845   & 29.098   \\
        -1.3 & 26.586   & 23.973   & 23.285   \\
        -1.2 & 20.944   & 19.354   & 18.735   \\
        -1.1 & 16.616   & 15.698   & 15.153   \\
        -1.0 & 13.273   & 12.784   & 12.320   \\
        -0.9 & 10.676   & 10.448   & 10.066   \\
        -0.8 & 8.644    & 8.563    & 8.264    \\
        -0.7 & 7.045    & 7.031    & 6.816    \\
        -0.6 & 5.779    & 5.779    & 5.648    \\
        -0.5 & 4.770    & 4.770    & 4.700    \\
        -0.4 & 3.962    & 3.962    & 3.928    \\
        -0.3 & 3.311    & 3.311    & 3.297    \\
        -0.2 & 2.783    & 2.783    & 2.779    \\
        -0.1 & 2.353    & 2.353    & 2.352    \\
         0.0 & 2.000    & 2.000    & 2.000    \\
        \hline
    \end{tabular}
    \caption{Critical values of $\alpha$ for different $\sigma$ values. The first column is the replica symmetric Parisi answer from equation \ref{acritsigmag0}. Column 2 a the replica-symmetric JPA, which we find gives precisely the same answer as the 1RSB Parisi ansatz. This answer differs significantly from the simple RS Parisi Ansatz. Column 3 is the 1RSB JPA value for $\acrit$, nearly the same as RS JPA.}
    \label{tab:critical_alpha}
\end{table}

The full RSB JPA is cumbersome to work with, so instead we consider the $k=1$ ansatz for simplicity. For $k=1$, we can write everything in terms of three quantities: $\gamma, q_0$ and $\tilde m_0$. The intensive action is
\begin{equation}
\begin{split}
    S&=\frac 12 \left(-\frac 1{\tilde m_{0}}\log \gamma+\left(\frac{1}{\tilde m_{0}}-1\right)\log\left(\gamma+(1-q_0)\tilde m_{0}\right)+\log \left(\gamma+(1-q_0)\tilde m_{0}+q_0\right)\right) \\
    &+\alpha\log \int_{-\infty}^\infty \exp\left(-\frac{h^2}{2q_0}\right)g(h)^{\frac{1}{\tilde{m}_{0}}}\frac{dh}{\sqrt{2\pi q_0}} \\
    &\textrm{where}\quad g(h)=\int_{-\infty}^\infty\exp\left(-\frac{\textrm{min}(h'-\sigma,0)^2\tilde m_{0}}{2\gamma}\right)\exp\left(-\frac{(h-h')^2}{2(1-q_0)}\right)\frac{dh'}{\sqrt{2\pi(1-q_0)}}.
\end{split}
\label{eq:1RSB}
\end{equation}
Now that we have Eq. (\ref{eq:1RSB}) for the action, we need to minimize it with respect to $\gamma,\tilde m_{0},$ and $q_0$. For $\sigma>0$, the action in Eq. (\ref{eq:1RSB}) is minimized when $\tilde m=1$ or $q_0=1$, corresponding to a replica symmetric matrix and giving us the same $\psat$ as in Section \ref{replicasymJPA}. However for any $\sigma<0$, the minimum occurs at an RSB solution. We find the transition is always second-order in $\alpha$. $\gamma$ and $\tilde m$ go as $(\alpha-\acrit)^{-1}$, meaning $\log \psat$ goes as $(\alpha-\acrit)^2$. 

\section{The Ground State Energy of the $p$-Spherical Model}\label{appendixgroundstatenergypspherical}
The JPA isn't just useful for calculation capacities and thresholds. Since ground state energies can also be phrased as constraint satisfaction, the JPA can be used to calculate the location and large deviations of the ground state of a disordered system. The technique is to take $\lim_{n\to 0} \overline{\zmic^n(E)}$, where $\zmic=\int  \delta(H(x)-E)d^Nx$ is the microcanonical partition function, the total volume of configuration space at energy $E$. The limit $\lim_{n\to0}\zmic^n(E)$ is 0 if there are no configurations of energy $E$, and 1 if there are. Thus the disorder average is precisely the probability that there are configurations at energy $E$. When it first dips below 1, we are below the quenched ground state energy.

This technique appears to work for a wide variety of disordered systems, but we will demonstrate it here for the $p$-spherical model \cite{CAG,Crisanti1993Spherical,Crisanti1995ThoulessAndersonPalmerAT,Cugliandolo1999RealTime,Cugliandolo1993Analytical,DAlessio2016From,PhysRevB.23.4661}. This is a simple statistical mechanical model with
\begin{equation}
    H(x)=\sum_{1\leq i_1\leq i_2...\leq i_p\leq N}J_{i_ii_2...i_p}x_{i_1}x_{i_2}...x_{i_p},
\end{equation}
that is a random $p$th order potential with coefficients chosen from an iid normal distribution with variance $\frac{p!}{N^{p-1}}$. The word ``spherical'' in the model's name refers to the spherical condition
\begin{equation}
    \sum_{i=1}^nx_i^2=N.
\end{equation}
Because the $J$s are all Gaussian random variables, the joint distribution of any collection of points will be jointly Gaussian with covariance
\begin{equation}
    \overline{H(x_\alpha)H(x_\beta)}=Nq_{\alpha\beta}^p.
\end{equation}
The replicated microcanonical partition function is
\begin{equation}
    \overline{\zmic^n(E)}=\overline{\int d^n x \int d^n\lambda \exp\left(\sum_{\alpha=1}^n\Lambda_\alpha(H(x_\alpha)-E)\right)}.
\end{equation}
We can perform the Gaussian integral to get
\begin{equation}
    \overline{\zmic^n(E)}=\int d^n x \int d^n\lambda \exp\left(\frac N2 \sum_{\alpha,\beta=1}^n\lambda^2 q_{\alpha\beta}^p-n\sum_{\alpha=1}^n\Lambda_{\alpha} E)\right).
\end{equation}
Integrating out the $x$s, we are left with
\begin{equation}
\log \overline{\zmic^n(E)}=\frac N2\log \det q+\frac N2 \sum_{\alpha=1}^n\lambda^2 q_{\alpha\beta}^p-n\Lambda E.
\end{equation}
We will assume a replica-symmetric JPA for the matrix $q$, and let $\Lambda$ scale as $L/n$. Under this ansatz
\begin{equation}
\begin{split}
    \log \det q=\log \frac{1+\gamma}{\gamma}\\
    \sum_{\alpha\beta}\lambda_{\alpha}\lambda_{\beta}q^p_{\alpha\beta}=pL^2\gamma
\end{split}
\end{equation}
This gives us the action
\begin{equation}
    \frac 1N\log \overline{\zmic^n(E)}=\frac 12\log \frac{1+\gamma}{\gamma}+\frac 12 (1+p\gamma)L^2-LE.
\end{equation}
We can solve for $L$ and get $L=\frac{E}{p\gamma}$, giving us
\begin{equation}
    \frac 1N\log \overline{\zmic^n(E)}=\min_\gamma\left(\frac 12\log \frac{1+\gamma}{\gamma}-\frac{E^2}{2(1+p\gamma )}\right).
\end{equation}
For, say $p=3$, this becomes negative at $E=-1.6569983$, precisely the point that replica thermodynamics predicts. Since the $p$-spherical model is a 1RSB-exact system, the fact that the RS JPA gets precisely the correct answer might lead to a conjecture that the $k$-level JPA somehow contains the same information as the $k+1$-level Parisi Ansatz. Unfortunately this doesn't seem to be borne out: for mixed $p$-spin models such as the ones in \cite{Auffinger_2018}, this conjecture appears to be false and the $k$th order JPA seems to contain the same information as $k$RSB Parisi. Working out a general principle behind these correspondences would be an interesting direction for future work.

\section{Discussion}

In this letter, we computed $\psat$ and $\acrit$ for the spherical perceptron using a novel generalization of the Parisi ansatz. For $\sigma > 0$, we found that the RS JPA accurately predicted $\psat$ and $\acrit$ in the large-$N$ limit. For $\sigma < 0$, we conjectured that a RSB JPA would yield the exact values of $\psat$ and $\acrit$, though we only explicitly evaluated the $1$RSB JPA. While the $1$RSP JPA produced results differing from the RS JPA, the discrepancy in $\acrit$ was less than $0.01\%$, indicating that the RS JPA is a good approximation for $\psat$ and $\acrit$.

One important collection of results has to do with the order of the transition in $\psat$. Once we introduce replica symmetry breaking in the JPA, we find that the transition is second order for all values of $\sigma$, meaning that $\log \psat\sim (\alpha-\acrit)^2$. This suggests that for respectably large values of $N$, one can add $O(\sqrt N)$ more constraints above the threshold before the satisfaction probability is significantly diminished.

By comparison, the transition appears to be first order in the context of the $p$-spherical model. Furthermore, since the RS JPA gets the location of the transition exactly right, and if the RS solution gives $\log P(E)~(E-E_{\textrm{crit}})^1$, then no higher-order ansatz can give a less negative value of $P$ such as $(E-E_{\textrm{crit}})^2$. Explaining why the $p$-spherical model has qualitatively different behavior than the spherical perceptron for both positive and negative $\sigma$ is an open question.

%\AH{add discussion about phase transition}

%Interestingly, the RS JPA computations agree exactly with $\acrit$ obtained from the $1$RSB Parisi ansatz. Combined with a similar result for the $p$-spherical model (see Appendix \ref{appendixgroundstatenergypspherical}), this leads us to conjecture that for computing $\acrit$, the $k$RSB JPA is equivalent to the $(k+1)$RSB Parisi ansatz. This would be analogous to the fact that a $k$RSB complexity calculation has the same sensitivity as a $(k+1)$RSB free energy calculation \cite{Annibale_2003,PhysRevE.68.061103}. Verifying or deriving this conjecture in a more general setting would be an interesting direction for future work.
Interestingly, in certainly circumstances, the JPA can be used to investigate $\punsat=1-\psat$ deep in the satisfiable phase. When the CSP corresponds to a convex optimization problem (as occurs in the spherical perceptron when $\sigma\geq 0$), the unsatisfiability of the CSP is dual to the satisfiability of some other CSP. By applying the JPA to this dual problem, we can calculate the exponentially small difference between $\psat$ and 1 for certain CSPs in the satisfiable regime.

Another promising avenue for investigation is the distribution of gaps, whose significance has been emphasized in Refs.~\cite{PhysRevLett.109.125502,Franz_2017}. These works primarily focused on the gap distribution of random CSPs in the UNSAT phase, providing insight into how close such elements were to satisfiability. In contrast, an analysis of $\lim_{n\rightarrow 0}\overline{Z^{n}}$ using the JPA conditions to the satisfiable instances of the CSP. This suggests that applying the methods of Refs.~\cite{PhysRevLett.109.125502,Franz_2017} in this context could yield the gap distribution for satisfiable elements, possibly giving some insight as to what happens when jammed systems are given time to anneal.

Finally, the tools of the JPA in hand, a natural next target is more structured satisfiability problems, such as classifying neural manifolds \cite{chung2016linear,chung2018classification}, or even data with covariances given by a colored Gaussian rather than the identity. 

%Regarding applications, as highlighted in the Introduction, the most concrete use of our results lies in identifying structure in datasets. A key technical challenge is developing efficient techniques to compute $F(Q)$ for general distributions.

%Finally, another application of the JPA is in the study of ground-state energies. As a proof of concept, we compute the ground-state energy of the $p$-spherical model in Appendix~\ref{appendixgroundstatenergypspherical}.

\subsection*{Acknowledgments}

We thank 
Jaron Kent-Dobias, Jonah Kudler-Flam and Vladimir Narovlansky for useful discussions. MW acknowledges DOE grant DE-SC0009988. AH is grateful to the Simons Foundation as well as the Edward and Kiyomi Baird Founders' Circle Member Recognition for their support. 

\bibliographystyle{apsrev4-1long}
\bibliography{main.bib}
\appendix
\section{Details of Finding $\acrit$}
The jammed phase corresponds precisely to cases where $\log \psat$ is negative. That is, for any value of $\alpha$ and $\sigma$ where the minimum of action \ref{eq:fullAction} over JPA matrices is negative, the spherical perceptron is jammed. Since the action is of the form 
\begin{equation}
    S=\textrm{determinant term}+\alpha*\textrm{integral term},
\end{equation}
the phase transition happens at
\begin{equation}
    \acrit=\min_{Q}\frac{-\textrm{determinant term}}{\textrm{integral term}}
\end{equation}
The special $Q$ where this minimum ratio happens is also the saddle-point matrix for $\alpha$ just above $\acrit$. As $\alpha$ increases further, the saddle-point can move.

For the replica symmetric JPA, the minimization process just boils down to minimizing over one number, $\gamma$. There is a small subtlety for $\sigma >\sstar\approx-0.6129...$. This minimum occurs at $\gamma=\infty$, but the minimum occurs at finite $\gamma$ for $\sigma<\sstar$. This means that the transition occurs in a region where both terms themselves (which scale as $\gamma^{-1}$) are zero for $\sigma>\sstar$, so $\frac{dS}{d\alpha}|_{\acrit}=0$. In contrast, $\frac{dS}{d\alpha}|_{\acrit}=0$ is negative and the transition appears first-order within the replica symmetric ansatz for $\sigma<\sstar$.

For the RSB JPA, the situation is more complex. Theoretically, we still only need to minimize the ratio of terms over all sets of $\gamma$, $q_{0}$ and $\tilde m$. The minimum now occurs when $\gamma, \tilde m$ go to infinity with their ratios fixed. This brings us to a similar situation to the case of the RS JPA with $\sigma>\sstar$, where both the determinant and integral terms of the action are zero at $\acrit$, so the transition is second-order.

\end{document}